\documentclass[12pt]{article}
\usepackage{amsmath}
\usepackage{sw20jart}
\usepackage{epic}
\usepackage{curves}

\newtheorem{theorem}{Theorem}[section]

\newtheorem{definition}[theorem]{Definition}

\newtheorem{problem}[theorem]{Problem}

\input{tcilatex}

\begin{document}

\author{Mark Korenblit \\
Department of Computer Science \\
Holon Institute of Technology, Israel\\
e-mail: korenblit@hit.ac.il\bigskip \\
Vadim E. Levit\\
Department of Computer Science and Mathematics\\
Ariel University, Israel\\
e-mail: levitv@ariel.ac.il }
\title{On the Optimal Representation of Algebraic Expressions of Fibonacci
Graphs}
\date{}
\maketitle

\begin{abstract}
The paper investigates relationship between algebraic expressions and
graphs. We consider a digraph called a Fibonacci graph which gives a generic
example of non-series-parallel graphs. Our intention in this paper is to
simplify the expressions of Fibonacci graphs and eventually find their
shortest representations. With that end in view, we describe the optimal
decomposition method for generating Fibonacci graph expressions that is
conjectured to provide these representations. Proof (or disproof) of this
conjecture is presented as an open problem.

\end{abstract}

\section{Introduction\label{intro}}

A \textit{graph }$G=(V,E)$ consists of a \textit{vertex set\ }$V$ and an 
\textit{edge set\ }$E$, where each edge corresponds to a pair $(v,w)$ of
vertices. If the edges are ordered pairs of vertices (i.e., the pair $(v,w)$
is different from the pair $(w,v)$), then we call the graph \textit{directed}
or\textit{\ digraph}; otherwise, we call it \textit{undirected}. If $(v,w)$
is an edge in a digraph, we say that $(v,w)$ \textit{leaves} vertex $v$ and 
\textit{enters} vertex $w$. A vertex in a digraph is a \textit{source} if no
edges enter it, and a \textit{sink} if no edges leave it.

A \textit{path} from vertex $v_{0}$ to vertex $v_{k}$ in a graph $G=(V,E)$
is a sequence of its vertices $\left[ v_{0},v_{1},v_{2},\ldots ,v_{k-1},v_{k}%
\right] $ such that $(v_{i-1},v_{i})\in E$ for $1\leq i\leq k$. $G$ is an 
\textit{acyclic graph} if there is no closed path $\left[ v_{0},v_{1},v_{2},%
\ldots ,v_{k},v_{0}\right] $ in $G$. A two-terminal directed acyclic graph (%
\textit{st-dag}) has only one source $s$ and only one sink $t$. In an
st-dag, every vertex lies on some path from $s$ to $t$.

A graph $G^{\prime }=(V^{\prime },E^{\prime })$ is a \textit{subgraph} of $%
G=(V,E)$ if $V^{\prime }\subseteq V$ and $E^{\prime }\subseteq E$. A graph $%
G $ is \textit{homeomorphic} to a graph $G^{\prime }$ (a \textit{homeomorph}
of $G^{\prime }$) if $G$ can be obtained by subdividing edges of $G^{\prime
} $ with new vertices.

We consider a \textit{labeled graph} which has labels attached to its edges.
Each path between the source and the sink (a \textit{sequential path}) in an
st-dag can be presented by a product of all edge labels of the path.

\begin{definition}
We define the sum of edge label products corresponding to all possible
sequential paths of an st-dag $G$ as the \textit{canonical expression }of $G$%
.
\end{definition}

\begin{definition}
An algebraic expression is called an \textit{st-dag expression} (a \textit{%
factoring of an st-dag} in \cite{BKS}) if it is algebraically equivalent to
the canonical expression of an st-dag. An st-dag expression consists of
terms (edge labels), the operators $+$ (disjoint union) and $\cdot $
(concatenation, also denoted by juxtaposition when no ambiguity arises), and
parentheses.
\end{definition}

\begin{definition}
We define the \textit{complexity of an algebraic expression} in two ways.
The complexity of an algebraic expression is (i) the total number of terms
in the expression including all their appearances (\textit{the first
complexity characteristic}) or (ii) the number of plus operators in the
expression (\textit{the second complexity characteristic}).
\end{definition}

We will denote the first and the second complexity characteristic of an
st-dag expression by $T(n)$ and $P(n)$, respectively, where $n$ is the
number of vertices in the graph.

\begin{definition}
An equivalent expression with the minimum complexity is called an \textit{%
optimal representation of the algebraic expression}.
\end{definition}

\begin{definition}
A \textit{series-parallel} \textit{graph} is defined recursively so that a
single edge is a series-parallel graph and a graph obtained by a parallel or
a series composition of series-parallel graphs is series-parallel.
\end{definition}

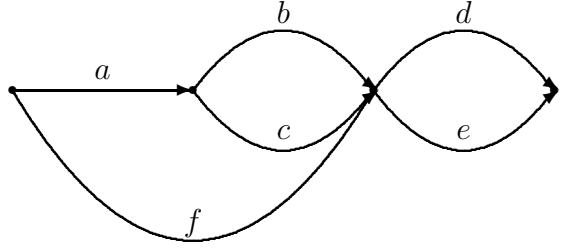
\begin{figure}[tbp]
\setlength{\unitlength}{0.8cm}
\par
\begin{picture}(5,4)(-3.5,0)\thicklines

\multiput(1,3)(3,0){4}{\circle*{0.15}}

\put(1,3){\vector(1,0){3}} \put(2.5,3.3){\makebox(0,0){$a$}}

\qbezier(4,3)(5.5,5)(7,3) \put(7.085,3){\vector(3,-2){0}}
\put(5.5,4.3){\makebox(0,0){$b$}}

\qbezier(4,3)(5.5,1)(7,3) \put(7.085,3){\vector(3,2){0}}
\put(5.5,2.3){\makebox(0,0){$c$}}

\qbezier(7,3)(8.5,5)(10,3) \put(10.085,3){\vector(3,-2){0}}
\put(8.5,4.3){\makebox(0,0){$d$}}

\qbezier(7,3)(8.5,1)(10,3) \put(10.085,3){\vector(3,2){0}}
\put(8.5,2.3){\makebox(0,0){$e$}}

\qbezier(1,3)(4,-2)(7,3) \put(7.085,3){\vector(4,3){0}}
\put(4,0.8){\makebox(0,0){$f$}}

\end{picture}
\caption{A series-parallel graph.}
\label{fig1}
\end{figure}

As shown in \cite{BKS} and \cite{KoL}, a series-parallel graph expression
has a representation in which each term appears only once. We proved in \cite%
{KoL} that this representation is an optimal representation of the
series-parallel graph expression from the perspective of the first
complexity characteristic. For example, the canonical expression of the
series-parallel graph presented in Figure \ref{fig1} is $%
abd+abe+acd+ace+fe+fd$. Since it is a series-parallel graph, the expression
can be reduced to $(a(b+c)+f)(d+e)$, where each term appears once.

\begin{definition}
A \textit{Fibonacci graph} \textit{(}$FG$\textit{) }\cite{GoP} has vertices $%
\{1,2,3,\ldots ,n\}$ and edges $\{\left( v,v+1\right) \mid v=1,2,\ldots
,n-1\}\cup \left\{ \left( v,v+2\right) \mid v=1,2,\ldots ,n-2\right\} $.
\end{definition}

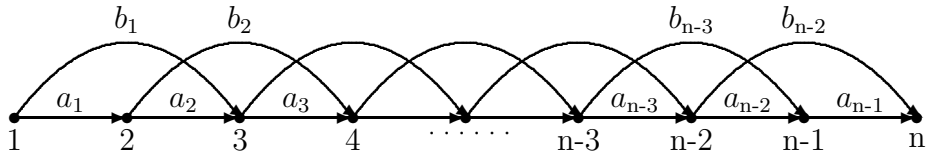
\begin{figure}[tbph]
\setlength{\unitlength}{1.0cm}
\par
\begin{picture}(5,2)(-0.9,-0.5)\thicklines

\multiput(0,0)(1.5,0){9}{\circle*{0.15}}

\put(0,-0.3){\makebox(0,0){1}}
\put(1.5,-0.3){\makebox(0,0){2}}
\put(3,-0.3){\makebox(0,0){3}}
\put(4.5,-0.3){\makebox(0,0){4}}
\put(7.5,-0.3){\makebox(0,0){n-3}}
\put(9,-0.3){\makebox(0,0){n-2}}
\put(10.5,-0.3){\makebox(0,0){n-1}}
\put(12,-0.3){\makebox(0,0){n}}

\multiput(0,0)(1.5,0){8}{\vector(1,0){1.5}}

\put(0.75,0.2){\makebox(0,0){$a_{1}$}}
\put(2.25,0.2){\makebox(0,0){$a_{2}$}}
\put(3.75,0.2){\makebox(0,0){$a_{3}$}}
\put(8.25,0.2){\makebox(0,0){$a_{\text{n-3}}$}}
\put(9.75,0.2){\makebox(0,0){$a_{\text{n-2}}$}}
\put(11.25,0.2){\makebox(0,0){$a_{\text{n-1}}$}}

\qbezier(0,0)(1.5,2)(3,0)
\qbezier(1.5,0)(3,2)(4.5,0)
\qbezier(3,0)(4.5,2)(6,0)
\qbezier(4.5,0)(6,2)(7.5,0)
\qbezier(6,0)(7.5,2)(9,0)
\qbezier(7.5,0)(9,2)(10.5,0)
\qbezier(9,0)(10.5,2)(12,0)

\multiput(3.085,0)(1.5,0){7}{\vector(3,-2){0}}

\put(1.5,1.3){\makebox(0,0){$b_{1}$}}
\put(3,1.3){\makebox(0,0){$b_{2}$}}
\put(9,1.3){\makebox(0,0){$b_{\text{n-3}}$}}
\put(10.5,1.3){\makebox(0,0){$b_{\text{n-2}}$}}

\multiput(5.55,-0.2)(0.2,0){6}{\circle*{0.02}}

\end{picture}
\caption{A Fibonacci graph.}
\label{fig2}
\end{figure}

As shown in \cite{Duf}, an st-dag is series-parallel if and only if it does
not contain a subgraph which is a homeomorph of the \textit{forbidden
subgraph} positioned between vertices $1$ and $4$ of the Fibonacci graph
illustrated in Figure \ref{fig2}. Thus, Fibonacci graphs are of interest as
\textquotedblright through\textquotedblright\ non-series-parallel st-dags.

Mutual relations between graphs and algebraic expressions are discussed in 
\cite{BKS}, \cite{GoM}, \cite{GMR}, \cite{KoL}, \cite{KoL1}, \cite{KoL2}, 
\cite{Mun1}, \cite{Mun2}, \cite{Nau}, \cite{SaW}, and other works.
Specifically, \cite{Mun1}, \cite{Mun2}, and \cite{SaW} consider the
correspondence between series-parallel graphs and read-once functions. A
Boolean function is defined as \textit{read-once} if it may be computed by
some formula in which no variable occurs more than once (\textit{read-once
formula}). On the other hand, a series-parallel graph expression can be
reduced to the representation in which each term appears only once. Hence,
such a representation of a series-parallel graph expression can be
considered as a read-once formula (Boolean operations are replaced by
arithmetic ones).

An expression of a homeomorph of the forbidden subgraph belonging to any
non-series-parallel st-dag has no representation in which each term appears
once. For example, consider the subgraph positioned between vertices $1$ and 
$4$ of the Fibonacci graph shown in Figure \ref{fig2}. Possible optimal
representations of its expression are $a_{1}\left( a_{2}a_{3}+b_{2}\right)
+b_{1}a_{3}$ or $\left( a_{1}a_{2}+b_{1}\right) a_{3}+a_{1}b_{2}$. For this
reason, an expression of a non-series-parallel st-dag can not be represented
as a read-once formula. However, for arbitrary functions, which are not
read-once, generating the optimum factored form is NP-complete \cite{Wan}.

Our intention is to simplify the expressions of Fibonacci graphs (we denote
them by $Ex(FG)$) and eventually find their optimal representations. The
last goal is an open problem. In this paper we survey a method which is
conjectured to provide an optimal representation for $Ex(FG)$.

\section{Preliminary Results}

The number of methods for generating Fibonacci graph expressions is
described in \cite{Kor}. Most of them derive representations with
complexities which increase exponentially as the number of the graph's
vertices increases.

Specifically, the \textit{sequential-paths method} is based directly on the
definition of an st-dag expression as the canonical expression of the
st-dag. For example, for a $9$-vertex Fibonacci graph, the corresponding
algebraic expression is 
\begin{eqnarray*}
&&a_{1}a_{2}a_{3}a_{4}a_{5}a_{6}a_{7}a_{8}+a_{1}a_{2}a_{3}a_{4}a_{5}a_{6}b_{7}+a_{1}a_{2}a_{3}a_{4}a_{5}b_{6}a_{8}+a_{1}a_{2}a_{3}a_{4}b_{5}a_{7}a_{8}+
\\
&&a_{1}a_{2}a_{3}a_{4}b_{5}b_{7}+a_{1}a_{2}a_{3}b_{4}a_{6}a_{7}a_{8}+a_{1}a_{2}a_{3}b_{4}a_{6}b_{7}+a_{1}a_{2}a_{3}b_{4}b_{6}a_{8}+
\\
&&a_{1}a_{2}b_{3}a_{5}a_{6}a_{7}a_{8}+a_{1}a_{2}b_{3}a_{5}a_{6}b_{7}+a_{1}a_{2}b_{3}a_{5}b_{6}a_{8}+a_{1}a_{2}b_{3}b_{5}a_{7}a_{8}+
\\
&&a_{1}a_{2}b_{3}b_{5}b_{7}+a_{1}b_{2}a_{4}a_{5}a_{6}a_{7}a_{8}+a_{1}b_{2}a_{4}a_{5}a_{6}b_{7}+a_{1}b_{2}a_{4}a_{5}b_{6}a_{8}+
\\
&&a_{1}b_{2}a_{4}b_{5}a_{7}a_{8}+a_{1}b_{2}a_{4}b_{5}b_{7}+a_{1}b_{2}b_{4}a_{6}a_{7}a_{8}+a_{1}b_{2}b_{4}a_{6}b_{7}+
\\
&&a_{1}b_{2}b_{4}b_{6}a_{8}+b_{1}a_{3}a_{4}a_{5}a_{6}a_{7}a_{8}+b_{1}a_{3}a_{4}a_{5}a_{6}b_{7}+b_{1}a_{3}a_{4}a_{5}b_{6}a_{8}+
\\
&&b_{1}a_{3}a_{4}b_{5}a_{7}a_{8}+b_{1}a_{3}a_{4}b_{5}b_{7}+b_{1}a_{3}b_{4}a_{6}a_{7}a_{8}+b_{1}a_{3}b_{4}a_{6}b_{7}+
\\
&&b_{1}a_{3}b_{4}b_{6}a_{8}+b_{1}b_{3}a_{5}a_{6}a_{7}a_{8}+b_{1}b_{3}a_{5}a_{6}b_{7}+b_{1}b_{3}a_{5}b_{6}a_{8}+
\\
&&b_{1}b_{3}b_{5}a_{7}a_{8}+b_{1}b_{3}b_{5}b_{7}.
\end{eqnarray*}%
It contains $201$ terms and $33$ plus operators.

\subsection{Decomposition method\label{sec_fg_dec}}

In \cite{KoL} we consider the \textit{decomposition method} which provides
an algorithm for constructing $Ex(FG)$ with polynomial complexity.

This method is based on revealing subgraphs in the initial graph. The
resulting expression is produced by a special composition of subexpressions
describing these subgraphs. 
\begin{figure}[tbph]
\setlength{\unitlength}{1.0cm}
\par
\begin{picture}(5,2)(-0.5,-0.4)\thicklines

\multiput(0,0)(1.5,0){10}{\circle*{0.15}}

\put(0,-0.3){\makebox(0,0){p}}
\put(1.5,-0.3){\makebox(0,0){p+1}}
\put(6,-0.3){\makebox(0,0){i-1}}
\put(7.4,-0.3){\makebox(0,0){i}}
\put(9,-0.3){\makebox(0,0){i+1}}
\put(12,-0.3){\makebox(0,0){q-1}}
\put(13.5,-0.3){\makebox(0,0){q}}

\multiput(0,0)(1.5,0){9}{\vector(1,0){1.5}}

\put(0.75,0.2){\makebox(0,0){$a_{\text{p}}$}}
\put(6.75,0.2){\makebox(0,0){$a_{\text{i-1}}$}}
\put(8.25,0.2){\makebox(0,0){$a_{\text{i}}$}}
\put(12.75,0.2){\makebox(0,0){$a_{\text{q-1}}$}}

\qbezier(0,0)(1.5,2)(3,0)
\qbezier(1.5,0)(3,2)(4.5,0)
\qbezier(3,0)(4.5,2)(6,0)
\qbezier(4.5,0)(6,2)(7.5,0)
\qbezier(6,0)(7.5,2)(9,0)
\qbezier(7.5,0)(9,2)(10.5,0)
\qbezier(9,0)(10.5,2)(12,0)
\qbezier(10.5,0)(12,2)(13.5,0)

\put(1.5,1.3){\makebox(0,0){$b_{\text{p}}$}}
\put(7.2,1.3){\makebox(0,0){$b_{\text{i-1}}$}}
\put(12,1.3){\makebox(0,0){$b_{\text{q-2}}$}}

\multiput(3.085,0)(1.5,0){8}{\vector(3,-2){0}}

\multiput(3.2,-0.2)(0.2,0){6}{\circle*{0.02}}
\multiput(10.05,-0.2)(0.2,0){6}{\circle*{0.02}}

\multiput(7.5,-0.7)(0,0.2){12}{\line(0,1){0.1}}

\end{picture}
\caption{Decomposition of a Fibonacci subgraph at vertex $i$.}
\label{fg_fig7}
\end{figure}
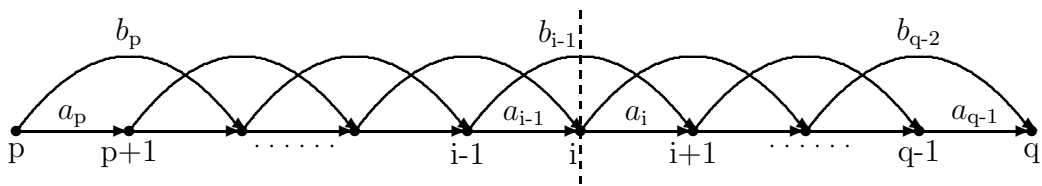

Consider the $n$-vertex $FG$ presented in Figure \ref{fig2}. Denote by $%
E(p,q)$ a subexpression related to its subgraph (which is an $FG$ as well)
having a source $p$ ($1\leq p\leq n$) and a sink $q$ ($1\leq q\leq n$, $%
q\geq p$). If $q-p\geq 2$, then we choose any \textit{decomposition vertex} $%
i$ ($p+1\leq i\leq q-1$) in a subgraph, and, in effect, split it at this
vertex (Figure \ref{fg_fig7}). Otherwise, we assign final values to $E(p,q)$%
. As follows from the structure of a Fibonacci graph, any path from vertex $%
p $ to vertex $q$ passes through vertex $i$ or avoids it via edge $b_{i-1}$.
Therefore, $E(p,q)$ can be generated by the following recursive procedure (%
\textit{decomposition procedure}):

\begin{enumerate}
\item \label{1}$\mathbf{case}$ $q=p:E(p,q)\leftarrow 1$

\item \label{2}$\mathbf{case}$ $q=p+1:E(p,q)\leftarrow a_{p}$

\item \label{3}$\mathbf{case}$ $q\geq p+2:\mathbf{choice}(p,q,i)$

\item \label{4}$\qquad \qquad \qquad \quad ~E(p,q)\leftarrow
E(p,i)E(i,q)+E(p,i-1)b_{i-1}E(i+1,q)$
\end{enumerate}

Lines \ref{1} and \ref{2} contain conditions of exit from the recursion. The
special case when a subgraph consists of a single vertex is considered in
line \ref{1}. It is clear that such a subgraph can be connected to other
subgraphs only serially. For this reason, it is accepted that its
subexpression is $1$, so that when it is multiplied by another
subexpression, the final result is not influenced. Line \ref{2} describes a
subgraph consisting of a single edge. The corresponding subexpression
consists of a single term equal to the edge label. The general case is
processed in lines \ref{3} and \ref{4}. The procedure, $\mathbf{choice}%
(p,q,i)$, in line \ref{3} chooses an arbitrary decomposition vertex $i$ on
the interval $(p,q)$ so that $p<i<q$. A current subgraph is decomposed into
four new subgraphs in line \ref{4}. Subgraphs described by subexpressions $%
E(p,i)$ and $E(i,q)$ include all paths from vertex $p$ to vertex $q$ passing
through vertex $i$. Subgraphs described by subexpressions $E(p,i-1)$ and $%
E(i+1,q)$ include all paths from vertex $p$ to vertex $q$ passing through
edge $b_{i-1}$.

$E(1,n)$ is the expression of the initial $n$-vertex $FG$ ($Ex\left(
FG\right) $). Hence, the decomposition procedure is initially invoked by
substituting parameters $1$ and $n$ instead of $p$ and $q$, respectively.

In \cite{KoL} we proved the following theorem that determines an optimal
location of the decomposition vertex $i$ in an arbitrary interval $(p,q)$ of
a Fibonacci graph from the perspective of the first complexity
characteristic.

\begin{theorem}
\label{th_fg-n/2}The representation with a minimum total number of terms
among all possible representations of $Ex(FG)$ derived by the decomposition
method is achieved if and only if in each recursive step $i$ is equal to $%
\frac{q+p}{2}$ for odd $q-p+1$ and to $\frac{q+p-1}{2}$ or $\frac{q+p+1}{2}$
for even $q-p+1$, i.e., when $i$ is a middle vertex of the interval $(p,q)$.
Such a decomposition method is called \textit{optimal}.
\end{theorem}

The following theorem for the second complexity characteristic is proven in 
\cite{Kor}.

\begin{theorem}
\label{th_fg-n/2_P}The representation with a minimum number of plus
operators among all possible representations of $Ex(FG)$ derived by the
decomposition method can be achieved by the optimal decomposition method.
\end{theorem}

It can be easily shown that for an $n$-vertex $FG$:\medskip

1. The total number of terms $T(n)$ in the expression $Ex(FG)$ derived by
the optimal decomposition method is defined recursively as follows: 
\begin{eqnarray*}
T(1) &=&0 \\
T(2) &=&1 \\
T(n) &=&T\left( \left\lceil \frac{n}{2}\right\rceil \right) +T\left(
\left\lfloor \frac{n}{2}\right\rfloor +1\right) +T\left( \left\lceil \frac{n%
}{2}\right\rceil -1\right) +T\left( \left\lfloor \frac{n}{2}\right\rfloor
\right) +1\quad (n>2).
\end{eqnarray*}

2. The number of plus operators $P(n)$ in the expression $Ex(FG)$ derived by
the optimal decomposition method is defined recursively as follows: 
\begin{eqnarray*}
P(1) &=&0 \\
P(2) &=&0 \\
P(n) &=&P\left( \left\lceil \frac{n}{2}\right\rceil \right) +P\left(
\left\lfloor \frac{n}{2}\right\rfloor +1\right) +P\left( \left\lceil \frac{n%
}{2}\right\rceil -1\right) +P\left( \left\lfloor \frac{n}{2}\right\rfloor
\right) +1\quad (n>2).
\end{eqnarray*}%
For large $n$%
\begin{equation*}
T(n)\approx 4T\left( \left\lceil \frac{n}{2}\right\rceil \right) +1
\end{equation*}%
and, by the \textit{master theorem} \cite{CLR}, $T(n)$ and $P(n)$ are $%
\Theta \left( n^{2}\right) $.

For $n=9$, the possible algebraic expression derived by the optimal
decomposition method is 
\begin{eqnarray*}
&&((a_{1}a_{2}+b_{1})(a_{3}a_{4}+b_{3})+a_{1}b_{2}a_{4})((a_{5}a_{6}+b_{5})(a_{7}a_{8}+b_{7})+a_{5}b_{6}a_{8})+
\\
&&(a_{1}(a_{2}a_{3}+b_{2})+b_{1}a_{3})b_{4}(a_{6}(a_{7}a_{8}+b_{7})+b_{6}a_{8}).
\end{eqnarray*}%
It contains $31$ terms and $11$ plus operators.

As shown in \cite{Kor}, the optimal decomposition method is not always the
only one that provides an expression for a Fibonacci graph with a minimum
number of plus operators. There exist \textit{special values of }$n$ when an 
$n$-vertex Fibonacci graph has several expressions with the same minimum
number of plus operators (among expressions derived by the decomposition
method). These special values are grouped as follows: 
\begin{equation*}
7,13\div 15,25\div 31,49\div 63,97\div 127,193\div 255,\ldots
\end{equation*}%
In the general view, they can be presented in the following way: 
\begin{eqnarray*}
n_{first_{\nu }} &\leq &n_{sp_{\nu }}\leq n_{last_{\nu }}, \\
n_{first_{1}} &=&n_{last_{1}}=7, \\
n_{first_{\nu }} &=&2n_{first_{\nu -1}}-1, \\
n_{last_{\nu }} &=&2n_{last_{\nu -1}}+1.
\end{eqnarray*}%
Here $\nu $ is a number of a group of special numbers; $n_{sp_{\nu }}$ is a
special number of the $\nu $-th group; $n_{first_{\nu }}$ and $n_{last_{\nu
}}$ are the first value and the last value, respectively, in the $\nu $-th
group. For all these values of $n$, not only the values of $i$ which are
mentioned in Theorem \ref{th_fg-n/2}, provide a minimum number of plus
operators in $Ex(FG)$.

For example, for $n=7$, the possible algebraic expression derived by the
optimal decomposition method ($i$ is equal to $4$ in the first recursive
step) is 
\begin{eqnarray*}
&&(a_{1}(a_{2}a_{3}+b_{4})+b_{1}a_{3})(a_{4}(a_{5}a_{6}+b_{5})+b_{4}a_{7})+
\\
&&(a_{1}a_{2}+b_{1})b_{3}(a_{5}a_{7}+b_{5}).
\end{eqnarray*}%
It contains $19$ terms and $7$ plus operators. For $i$ chosen equal to $3$
in the first recursive step, the possible expression is 
\begin{eqnarray*}
&&(a_{1}a_{2}+b_{1})((a_{3}a_{4}+b_{3})(a_{5}a_{6}+b_{5})+a_{3}b_{4}a_{6})+
\\
&&a_{1}b_{2}(a_{4}(a_{5}a_{6}+b_{5})+b_{4}a_{6}).
\end{eqnarray*}%
This expression contains $20$ terms but the number of its plus operators is
also equal to $7$.

\subsection{Generalized decomposition (GD) method\label{sec_GD}}

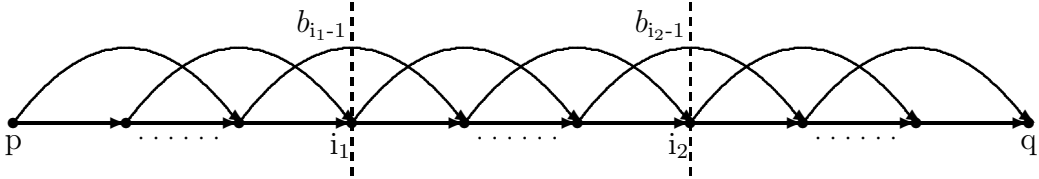
\begin{figure}[tbph]
\setlength{\unitlength}{1.0cm}
\par
\begin{picture}(5,2)(-0.5,-0.4)\thicklines

\multiput(0,0)(1.5,0){10}{\circle*{0.15}}

\put(0,-0.3){\makebox(0,0){p}}
\put(4.35,-0.3){\makebox(0,0){i$_{\text{1}}$}}
\put(8.85,-0.3){\makebox(0,0){i$_{\text{2}}$}}
\put(13.5,-0.3){\makebox(0,0){q}}

\multiput(0,0)(1.5,0){9}{\vector(1,0){1.5}}

\qbezier(0,0)(1.5,2)(3,0)
\qbezier(1.5,0)(3,2)(4.5,0)
\qbezier(3,0)(4.5,2)(6,0)
\qbezier(4.5,0)(6,2)(7.5,0)
\qbezier(6,0)(7.5,2)(9,0)
\qbezier(7.5,0)(9,2)(10.5,0)
\qbezier(9,0)(10.5,2)(12,0)
\qbezier(10.5,0)(12,2)(13.5,0)

\put(4.1,1.3){\makebox(0,0){$b_{\text{i}_{\text{1}}\text{-1}}$}}
\put(8.6,1.3){\makebox(0,0){$b_{\text{i}_{\text{2}}\text{-1}}$}}

\multiput(3.085,0)(1.5,0){8}{\vector(3,-2){0}}

\multiput(1.7,-0.2)(0.2,0){6}{\circle*{0.02}}
\multiput(6.2,-0.2)(0.2,0){6}{\circle*{0.02}}
\multiput(10.7,-0.2)(0.2,0){6}{\circle*{0.02}}

\multiput(4.5,-0.7)(0,0.2){12}{\line(0,1){0.1}}
\multiput(9,-0.7)(0,0.2){12}{\line(0,1){0.1}}

\end{picture}
\caption{Decomposition of a Fibonacci subgraph at vertices $i_{1}$ and $%
i_{2} $.}
\label{fg_fig8}
\end{figure}

As follows from the previous section, the decomposition method is based on
splitting a Fibonacci graph in each recursive step into two parts via
decomposition vertex $i$ and edge $b_{i-1}$. The GD method entails splitting
a Fibonacci graph in each recursive step into an arbitrary number of parts
(we will denote this number by $m$) via \textit{decomposition vertices} $%
i_{1},i_{2},\ldots ,i_{m-1}$ and edges $b_{i_{1}-1},b_{i_{2}-1},\ldots
,b_{i_{m-1}-1}$, respectively. An example for $m=3$ is illustrated in Figure %
\ref{fg_fig8}.

In all cases when $m>2$, the decomposition procedure used in the previous
section is transformed to the more complex form. Specifically, for $m=3$,
the general line of the new decomposition procedure, corresponding to line %
\ref{4} of the decomposition procedure with $m=2$ is presented as: 
\begin{eqnarray*}
~E(p,q) &\leftarrow &E(p,i_{1})E(i_{1},i_{2})E(i_{2},q)+ \\
&&E(p,i_{1}-1)b_{i_{1}-1}E(i_{1}+1,i_{2})E(i_{2},q)+ \\
&&E(p,i_{1})E(i_{1},i_{2}-1)b_{i_{2}-1}E(i_{2}+1,q)+ \\
&&E(p,i_{1}-1)b_{i_{1}-1}E(i_{1}+1,i_{2}-1)b_{i_{2}-1}E(i_{2}+1,q).
\end{eqnarray*}%
The sum above consists of four parts, with each part including three
subexpressions corresponding to the three parts of a split subgraph. Hence,
a current subgraph is decomposed into twelve new subgraphs.

Suppose that a Fibonacci graph is split into approximately equal parts in
each recursive step (distances between decomposition vertices are equal or
approximately equal). It will be the \textit{uniform GD method}.

The following theorem is proven in \cite{KoL1}.

\begin{theorem}
\label{th_gd}For an $n$-vertex $FG$, both the total number of terms $T(n)$
and the number of plus operators $P(n)$ in the expression $Ex(FG)$ derived
by the uniform GD method (the $FG$ is split into $m$ parts) are $O\left(
n^{1+\log _{m}2^{m-1}}\right) $.
\end{theorem}

As follows from Theorem \ref{th_gd}, $T(n)$ and $P(n)$ reach the minimum
complexity among $2\leq m\leq n-1$ when $m=2$. Substituting $2$ for $m$
gives $O\left( n^{2}\right) $ (we have the optimal decomposition method in
this case). Further, the complexity increases with the increase in $m$. For
example, we have $O\left( n^{1+\log _{3}4}\right) $ for $m=3$, $O\left(
n^{2.5}\right) $ for $m=4$, etc. In the extreme case, when $m=n-1$, all
inner vertices (from $2$ to $n-1$) of an $n$-vertex $FG$ are decomposition
vertices. The single recursive step is executed in this case, and all
revealed subgraphs are individual edges (labeled $a$ with an index)
connected by additional edges (labeled $b$ with an index). That is, in this
instance, the uniform GD method is reduced to the sequential-paths method.
Substituting $n-1$ for $m$ gives 
\begin{equation*}
O\left( n^{1+\log _{n-1}2^{n-2}}\right) >O\left( n^{1+\log
_{n}2^{n-2}}\right) =O\left( 2^{n-2}n\right) .
\end{equation*}

\section{Open Problems}

We conjecture that the optimal decomposition method provides an optimal
representation (for both our complexity characteristics) of an algebraic
expression related to a Fibonacci graph. The results obtained in section \ref%
{sec_GD} do not contradict this conjecture. At least,the optimal
decomposition method is the best one among uniform GD methods
(asymptotically).

However, we did not prove that splitting a Fibonacci graph into
approximately equal $m$ parts gives the optimal result for arbitrary $m$ (as
in Theorems \ref{th_fg-n/2} and \ref{th_fg-n/2_P} for $m=2$). Besides, the
GD method entails splitting a Fibonacci graph into the same number of parts
in each recursive step. One further generalization of the method assigns to
any subgraph its own number of decomposition vertices. Finally, there exist
representations that are obtained through algorithms which are not
appropriate to any generalized decomposition method.

Thus, we have the following open problems.

\begin{problem}
Prove (or disprove) that the optimal decomposition method is the only one
that provides an optimal representation of an algebraic expression related
to a Fibonacci graph from the perspective of the first complexity
characteristic.
\end{problem}

\begin{problem}
Prove (or disprove) that the optimal decomposition method provides an
optimal representation of an algebraic expression related to a Fibonacci
graph from the perspective of the second complexity characteristic.
\end{problem}

\end{document}